# Lattice Dynamics and Thermal Expansion Behavior in Metal Cyanides, MCN (M=Cu, Ag, Au): Neutron Inelastic Scattering and First Principles Calculations

M. K. Gupta[1], Baltej Singh[1], R. Mittal[1], S. Rols[2] and S. L. Chaplot[1]

[1]*Solid State Physics Division, Bhabha Atomic Research Centre, Mumbai 400085*

[2]*Institut Laue-Langevin, 71 Avenue des Martyrs, CS 20156, 38042 Grenoble Cedex 9, France*

We report measurement of temperature dependence of phonon spectra in quasi one-dimensional metal cyanides MCN (M=Cu, Ag and Au). *Ab-initio* lattice dynamics calculations have been performed to interpret the phonon-spectra as well as to understand the anamolous thermal expansion behavior in these compounds. We bring out the differences in the phonon mode behavior to explain the differences in the thermal expansion behavior among the three compounds. The chain-sliding modes are found to contribute maximum to the negative thermal expansion along 'c' axis in the Cu- and Ag- compounds, while the same modes contribute to positive thermal expansion in the Au- compound. Several low energy transverse modes lead to positive thermal expansion along 'a' and 'b' axis in all the compounds. The calculated elastic constants and Born effective charges are correlated with the difference in nature of bonding among these metal cyanides.

# I. INTRODUCTION

The phenomenon of large negative thermal expansion (NTE) in $ZrW_2O_8$ over a wide range of temperatures has lead to extensive research[1-17] in this area since last two decades. In fact, the phenomenon was observed long back in water. The polyhedral framework compounds with large open structure are mostly found to exhibit this phenomenon e.g $ZrV_2O_7$, $TaV_2O_5$, $Sc_2(MoO_4)_3$, $HfV_2O_7$ etc. The discovery of colossal thermal expansion behavior in metal cyanides compounds[15, 18, 19] has further accelerated the research in this field. The discovery of NTE has lead to industrial applications of these compounds in various areas like fiber optics, coatings, electronics and mirror substrates to tooth fillings etc.

Most of the compounds exhibiting negative thermal expansion behavior consist of rigid polyhedral units around metal ions. The polyhedral units are mutually connected via terminal oxygens. The terminal oxygen plays an important role in governing the thermal expansion behavior in such compounds. $ZrW_2O_8$ is very popular compound in this category. It shows isotropic negative thermal expansion behavior from 0.3 K to 1050 K. Its structure remains cubic up to 1050 K, however there is an order-disorder transition around 450 K. The low energy transverse phonon modes in the compound are found to be very anharmonic in nature, which led to transverse displacement of oxygen and causes NTE in $ZrW_2O_8$[20-23]. The other frame work compounds such as $ZrV_2O_7$, $HfV_2O_7$ also show similar behavior although there structures are different than that of $ZrW_2O_8$. A number of experimental and theoretical studies have been performed on such compounds. These studies suggest that the transverse vibration of oxygen and distortions of polyhedron units are mainly responsible for negative thermal expansion.

In the case of $M_2O$ (M=Ag, Cu and Au)[24-26] compounds, the metal ion M acts as terminal atom to connect $M_4O$ polyhedral and plays important role in NTE behavior. The difference in the thermal expansion behaviour in these compounds mainly arises due to the difference in nature of bonding. The phonon calculations show that NTE in oxygen mediated compound is mainly supported by the structure *i.e.* open structure. However in compounds where metal atoms play the role of terminal entity, the nature of bonding between metal atom and oxygen is also found to be important in governing the thermal expansion behavior.

Recently metal cyanides have gained attention. $Zn(CN)_2$ was the first compound in this category[27]. The word "colossal" is used to emphasis the huge thermal expansion behavior of the compound. The structure consists of tetrahedral framework of Zn connected with four cyanide units.

The tetrahedral units are connected through CN, which provide the flexibility to bend the tetrahedra and results in NTE behaviour. The isostrucral cyanide Cd(CN)$_2$ also exhibit NTE behavior. The magnitude of negative thermal expansion coefficient is found to be larger[17] in comparison to Zn(CN)$_2$. The first principles calculations have been done to understand the difference in magnitude of negative thermal expansion coefficient in these two compounds. The calculations shows that low energy phonons in Cd(CN)$_2$ are more sensitive to pressure than in Zn(CN)$_2$. NTE has been explained in terms of rotations, translations and deformations of M(C/N)$_4$ coordinated tetrahedra[28, 29].

Recently a number of prussian blue analogue were also found to be exhibit NTE, where the feature is attributed from metollopholic interactions[19,30,31]. The studies have been done to see the impact of metallophilicity on colossal positive and negative thermal expansion in a series of isostructural dicyanometallate coordination polymers[32,33]. The cyanide compounds are found to exhibit the phenomenon of negative linear compressibility. The c-axis in Ag$_3$Co(CN)$_6$ is found to show NTE as well as negative linear compressibility[9,18,34-37] behavior. KMn[Ag(CN)$_2$]$_3$ shows extreme negative compressibility mediated by selective soft-mode frustration[38]. All these compounds show NTE behavior due to their polyhedral dynamics and include rotation and distortion. The compounds consist of polyhedra units connected via C≡N, and are known to provide more flexibility in comparison to those where polyhedrals are connected via oxygens. It seems the structure consisting of -C≡N- units provide much flexibility for the bending motion, which is found to be one of the major cause of NTE behavior in cyanides.

The thermal expansion behavior in low dimension MCN (M=Cu, Ag and Au) compounds shows anisotropic thermal expansion behavior. The structure of the compounds has been determined by various groups[10,39-44]. X-ray powder diffraction measurements have been performed[39] over a temperature range of 90-490 K. The structure of the cyanides is chain like and resembles a quasi one dimension structure. These chains consist of C≡N units connected via metal ions (M-C≡N-M). The structure seems to be simple, however the compounds shows C/N disorder along the chain in terms of random flipping of C/N sequence. The higher dimension cyanides like Cd(CN)$_2$, Ni(CN)$_2$ and Zn(CN)$_2$ are also known to show C/N disorder behavior.

CuCN crystallizes in two different structures named as low temperature and high temperature phase at ambient condition depends on the method of synthesis. The low temperature (orthorhombic, C222$_1$) and high temperature (hexagonal, R3m) phase of CuCN in the manuscript are termed by LT-CuCN and HT-CuCN, respectively. The low temperature phase is a modulated structure of the high temperature phase. The modulation in LT-CuCN[44] is observed from previous neutron diffraction study. The structure of LT-CuCN consists of long Cu-C≡N-Cu modulated chains, each containing five crystallographically distinct Cu atoms, which form a wave consisting of nine CuCN units.

At ambient condition, AgCN and HT-CuCN crystallizes in hexagonal R3m (space group no. 160) cell. However AuCN crystallizes in P6mm (space group no. 183) structure. The unit cell of AgCN and HT-CuCN consist of three formula units. However the structure of AuCN consists of single formula unit. All the three compounds have three atoms in their primitive unit cell. The crystal structures as shown in Fig. 1 indicates that in AuCN all chains (M-C≡N-M) are parallel along to c-axis while in HT-CuCN and AgCN the adjacent parallel chains are shifted by an amount of c/3 along c-axis. Earlier reverse Monte Carlo simulations of the diffraction data have been performed to understand the local structure of these cyanides[39]. The buckling in the M-C≡N-M chains is found[39] to increase with temperature. The magnitude of buckling is governed by nature of bonding between metal ions and C≡N unit. The analysis suggests that HT-CuCN have large distortion perpendicular to the chain direction. Similar behavior is also observed in AgCN and AuCN; however, the magnitude of such distortion is very small in AuCN. The magnitude of distortion in all three compounds increases with temperature and found to be correlated[39] with the thermal expansion coefficient along the chain direction ($\alpha_c$). The thermal expansion coefficient is positive in the a-b plane, however large NTE is found along the chain. The coefficient of negative thermal expansion along the chain direction for HT-CuCN, LT-CuCN, AgCN and AuCN is $-27.9\times10^{-6}$ K$^{-1}$, $-53.8\times10^{-6}$ K$^{-1}$, $-14.8\times10^{-6}$ K$^{-1}$ and $-6.9\times10^{-6}$ K$^{-1}$ respectively[39, 44].

Raman and infrared measurements have also been done[43] on MCN compounds. These measurements are limited to zone centre, hence they do not provide complete information about dynamics of the compounds. Here we present the temperature dependent inelastic scattering measurements on these cyanides. The measured spectra have contributions from all phonon modes from entire Brillouin zone. Our studies provide vibrational properties of these cyanides and the analysis of vibrational spectra using ab-initio phonon calculations is useful to understand the thermal expansion behavior of these cyanides.

## II. EXPERIMENTAL DETAILS

The polycrystalline sample of AgCN, AuCN and LT-CuCN were purchased from Sigma Aldrich. The HT-CuCN was prepared by heating Aldrich supplied LT-CuCN for two hours under vacuum at 325°C. The inelastic neutron scattering experiments on MCN (M=Cu, Ag and Au) were carried out using the IN4C spectrometers at the Institut Laue Langevin (ILL), France. The spectrometer is based on the time-of-flight technique and is equipped with a large detector bank covering a wide range of about $10^o$ to $110^o$ of scattering angle. The polycrystalline samples of MCN were prepared by the solid state reaction method. The inelastic neutron scattering measurements were performed at three temperatures from 150 K, 240 K and 310 K. About 1 cc of polycrystalline samples of MCN have been used for the measurements. The low temperature measurements were performed using a helium cryostat. For these measurements we have used an incident neutron wavelength of 2.4 Å (14.2 meV) in neutron energy gain setup. In the incoherent one-phonon approximation, the measured scattering function $S(Q,E)$, as observed in the neutron experiments, is related[45-47] to the phonon density of states $g^{(n)}(E)$ as follows:

$$g^{(n)}(E) = A \left< \frac{e^{2W(Q)}}{Q^2} \frac{E}{n(E,T) + \frac{1}{2} \pm \frac{1}{2}} S(Q,E) \right> \qquad (1)$$

$$g^n(E) = B \sum_k \left\{ \frac{4\pi b_k^2}{m_k} \right\} g_k(E) \qquad (2)$$

where the + or − signs correspond to energy loss or gain of the neutrons respectively and where $n(E,T) = \left[ \exp(E/k_B T) - 1 \right]^{-1}$. $A$ and $B$ are normalization constants and $b_k$, $m_k$, and $g_k(E)$ are, respectively, the neutron scattering length, mass, and partial density of states of the $k^{th}$ atom in the unit cell. The quantity between < > represents suitable average over all $Q$ values at a given energy. $2W(Q)$ is the Debye-Waller factor averaged over all the atoms. The weighting factors $\frac{4\pi b_k^2}{m_k}$ for various atoms in the units of barns/amu are: 0.1264, 0.04626, 0.03944, 0.4625 and 0.8221 for Cu, Ag, Au, C and N respectively. The values of neutron scattering lengths for various atoms can be found from Ref. [48-50].

### III. COMPUTATIONAL DETAILS

The ab-initio phonon spectra can be calculated using either from supercell method or density functional perturbation theory (DFPT)[51] method. We have used the former approach of supercell method to calculate the phonon frequencies. The PHONON software[52] is used to generate supercell with various (±x, ±y, ±z) atomic displacement patterns. The Hellman-Feynman forces on the atoms in the super cell have been calculated using density functional theory as implemented in the VASP software[53]. The phonon calculations for HT-CuCN, AgCN and AuCN are performed considering the periodic lattice model using the experimental structure parameters as given in Table I. The model is an approximation of the real situation where we have neglected the C/N disorder. The low temperature phase of CuCN is a modulated structure of the high temperature phase. The required supercell to perform the calculations makes it computationally very expensive. Our interest is to understand the differences in thermal expansion behavior in terms of vibration, elastic constants and nature of bonding in these quasi one-dimensional metal cyanide systems, hence we have performed theoretical analysis on linear systems only.

There are 3 atoms in the primitive unit cell of HT-CuCN, AgCN (R3m) and AuCN (P6mm) phase, which gives 9 degree of freedom. The 9 displacement patterns are required to compute the phonon frequencies. For accurate force calculations we displaced the atoms in both the direction (±x,±y,±z) hence the number of displacements are double (18). The energy cutoff is 580 eV and a 8 ×8 × 8 k point mess have been used to obtain energy convergence in total energy of the order of meV, which is sufficient to obtain the required accuracy in phonon energies. The Monkhorst Pack method is used for k point generation[54]. The exchange-correlation contributions were approached within PBE generalized gradient approximation (GGA)[55]. The convergence criteria for the total energy and ionic forces were set to $10^{-8}$ eV and $10^{-5}$ eV Å$^{-1}$, respectively. The phonon spectra have been calculated in partially relaxed configuration. In partially relaxed only atomic coordinates are relaxed at fixed lattice parameter obtained from neutron diffraction data at 10 K and 310 K[39,42,44,56].

## IV. RESULTS AND DISCUSSION

### A. Temperature Dependence of Phonon Spectra

We have measured (Figs. 2 and 3) the inelastic neutron spectra of MCN (M=Cu, Ag and Au) at 150 K, 240 K and 310 K. As mentioned above the measurements are carried out in the energy gain mode which has allowed us to measure only the external modes. The C≡N stretching modes appear around 250 meV and would not be possible to measure due to the paucity of high energy phonons in the temperature range of the measurements. The phonon spectra of AgCN show peaks (Fig. 2) at about 4 meV, 16 meV, 36 meV and 55 meV. Further we observe the peak about 36 meV soften with temperature. However the peak at about 55 meV becomes more diffusive as temperature increases from 150 to 310 K. The intensity of these peaks decrease as the temperature rises. The increase in temperature will enhance the vibrational mean square amplitude of atoms; hence the Debye Waller factor would in turn reduces the intensity of the peaks. Also the C/N disorder is known to increase[39] with temperature which will further reduce the sharp features in the phonon spectra.

Further the measured spectra of AuCN show (Figs. 2 and 3) much broad features rather than sharp peaks as seen in spectra of other compounds. The lowest peak is about 4 meV and the others peaks are observed around 20 meV, 35 meV, 50 meV and 75 meV. We could not observe any significant softening of phonon modes in AuCN with temperature.

However in case of HT-CuCN we observe (Figs. 2 and 3) that the lowest peak in the phonon spectra is at about 7 meV and other peaks are around 20 meV, 45 meV and 70 meV. The lowest energy modes are shifted to high energies in comparison to AgCN. This could be partly due to the difference of mass of Cu (63.54 amu) and Ag (107.87 amu). We observe significant softening with temperature for phonon modes around 45 meV. The magnitude of softening in HT-CuCN is larger than that in AgCN.

In Fig. 3(b) we have shown the neutron inelastic scattering spectra measured at 150 K for high temperature and low temperature phases of CuCN. We find that the peak at around 20 meV in LT-CuCN seems to be broader in comparison to that in HT-CuCN, while at 300 K (Fig. 3(a)) the width of peaks in both the compounds seems to be same. The larger width at low temperature in LT-CuCN may be due to the fact that low temperature phase is a modulated structure of the high temperature phase. It seems at higher temperature the effect due to the CN disorder and anharmonicity dominate and the inelastic spectra as measured in both the phases appear similar.

**B. Calculated Phonon Spectra and Elastic Constants**

The crystal structure of all three metal cyanides is known to show C/N disorder [39]. The C-N are randomly oriented along the c direction. The ideal structure of HT-CuCN and AgCN consists of chains of -M-CN-M- along c-axes. The ab-initio phonon calculations are carried out considering the ordered structure of these compounds. The phonon spectra have been calculated at fixed lattice parameters corresponding to experimental structures[39, 42, 56] at 10 K and 310 K.

We have also calculated the phonon dispersion (Fig 5) relation of all three compounds along various high symmetry directions namely [100], [001] and [110]. We find that in all the three compounds transverse acoustic modes along [001] are unstable. The mode involves transverse motion of C and N atoms in the a-b plane. The C/N disorder in the compounds might be responsible for stability of the crystal.

The comparison between the experimental and calculated phonon spectra is shown in Fig. 4. The calculated spectra are able to reproduce all the major features of the observed spectra. The structural disorder could lead to a variation of the M-C, M-N and C-N bond lengths, which would in turn broadens the peaks as observed in the experimental spectra. This might be one of the reasons for difference in the calculated and experimental spectra of MCN. We notice that for HT-CuCN and AgCN elastic instability is observed along [100] and [110]. However, for AuCN these modes are found to be stable. The slopes of the transverse acoustic phonon branches (Fig. 5) are very low. Hence small errors in the calculation of phonon energies may result in large errors in the calculated elastic constants. So the elastic constant of MCN are calculated (TABLE II) using the symmetry-general least squares method[57] as implemented in VASP5.2 and were derived from the strain−stress relationships obtained from six finite distortions of the lattice. The calculated elastic moduli include contributions of distortions with rigid ions and ionic relaxations. The elastic constants $C_{11}$ and $C_{33}$ are related to the longitudinal phonons polarized along x and z axis. It can be seen that there is a large difference in the values of the $C_{11}$ and $C_{33}$ elastic constants in all the compounds. This indicates large difference in the nature of bonding in a-b plane and along c-axis. This is in agreement with the analysis of experimental diffraction data which also shows strong one dimensional nature of these compounds. The values of $C_{33}$ for HT-CuCN, AgCN and AuCN are 536 GPa, 387 GPa and 755 GPa respectively. Large value of $C_{33}$ in AuCN (~755 GPa) in comparison to the other two compounds indicates that bonding between the atoms of -Au-CN-Au- chains is much stronger in comparison to that in Ag and Cu compounds.

The $C_{66}$ elastic constant in all three compounds is very small. All these suggest that CuCN and AgCN are close to instability in plane against shear strain. However the AuCN shows significant stability against the shear strain. On increasing temperature, the magnitude of strain arising due to the vibrational amplitude of atoms perpendicular to chain will depend on the bond strength of -M-CN-M-. The calculated elastic constants as given in TABLE II indicates that nature of bonding in AuCN is strongest among all the three cyanides. This is consistence with the reverse Monte Carlo analysis of the diffraction data, which indicates that AuCN does not show any shear distortion even up to 450 K; however, significant distortion is observed in HT-CuCN and AgCN.

**C. Partial Phonon Density of States**

The partial density of states provides the contributions of the individual atoms to the total phonon spectra. We have calculated (Fig. 6) the partial density of states by projecting the eigenvector on different atoms. The contribution from M (Cu, Ag and Au) atoms is spread up to 35 meV; however, it is most significant contribution only below 10 meV. The C and N atoms contribute in the entire energy range up to 280 meV. We observed band gap in the phonon spectra from 80 meV to 280 meV. The CN stretching modes are at around 280 meV. The low energy peak in the partial density of states of Cu (63.54 amu), Ag (107.87 amu) and Au (197.97 amu) are at 7 meV, 5.2 meV and 5.2 meV respectively. The shift in the peak position is partly due to the mass renormalization. It should be noted that volume per primitive cell of HT-CuCN and AuCN compounds is nearly the same at 49.41 Å$^3$ and 49.82 Å$^3$ respectively. The lowest energy peak in the Au compound does not follow the mass effect. This indicates that the nature of bonding for the AuCN (P6mm) is stronger in comparison to HT-CuCN and AgCN (both in R3m). The difference in ionic radii of Cu (0.73 Å), and Au (1.37 Å) along with the similarity in volume per primitive cell of these compounds further supports the idea of difference in nature of bonding.

The partial contributions due to C and N atoms in the HT-CuCN (49.41 Å$^3$) and AgCN (53.48 Å$^3$) in the external mode region (below 80 meV) are up to 75 meV and 62 meV respectively. As expected the difference in the energy range of the external modes in the two compounds seems to follow the volume effect. The volume of the primitive unit cell of AuCN is 49.82 Å$^3$. We find that external modes in AuCN extend up to 80 meV. Comparisons of the energy range of the external modes of the three compounds suggest that force constants are stiffer in AuCN in comparison to other two compounds. The calculated C-N bond lengths (Table IV) are 1.174 Å, 1.169 Å and 1.163 Å in HT-CuCN, AgCN and AuCN respectively. As expected (Fig. 6) the energies of the phonon modes in the internal mode region simply follow the considerations due to change in C-N bond lengths. From the above analysis we can conclude that the nature of bonding along -Au-CN-Au- is stronger than that in -Ag-CN-Ag- and -Cu-CN-Cu-. This is consistent with the calculated Born effective charges as discussed in Section IVD.

In Fig 7 we have calculated the partial density of states at crystal structures corresponding to the structures at 10 K and 310 K. It can be seen that significant change in the phonon spectra with temperature is observed in HT-CuCN and AgCN, however in AuCN the change with temperature is not significant. The contributions due to the C and N atoms show large change around 45 meV and 75 meV in HT-CuCN and 35 meV and 60 meV in AgCN. In AuCN the spectral change of C and N is observed only around 80 meV. Interestingly in case of HT-CuCN the phonons around 35 meV soften with temperature in contrast to 75 meV which become harder. Similar behavior is also observed in AgCN for phonons of energy around 35 meV and 60 meV, however the magnitude of anharmonicity seems to be different. The phonon of energies around 80 meV in AuCN harden with increase in temperature. The contribution of metal ions in phonon spectra are limited to low energies and the observed changes with temperature are not very significant.

**D. Born Effective Charges**

The computed Born effective charges in all three compounds are listed in Table III. We observe that the values of the charge of carbon, nitrogen and M(Cu, Ag and Au) atoms in all three compound are different and anisotropic. This anisotropic behavior of the Born effective charge suggest difference in nature of bonding along a and c-axis. We find that for M atoms the values of the Born-effective charges along a-axis is large in comparison to that along c-axis. However, for C and N atoms this trend is reverse. It is interesting to note that for Cu and Ag compounds, the magnitude of Born effective charges along the chain (c-axis) has small finite value however in Au compound the value is zero. This suggests a large difference in nature of bonding in along M-CN-M chain among various MCN. This could be due to the difference in electronegativity of Cu(1.9), Ag(1.93) and Au(2.54) atoms. The zero magnitude of Born effective charge for Au compound along c-axis means that bonding along the chain may be either metallic or covalent. However, MCN are known to be insulator, hence the bonding between Au and CN may be covalent in nature.

Further for AuCN (Table II) the magnitude of $C_{33}$ and $C_{44}$ elastic constants, which are related to the longitudinal and transverse phonon frequencies along c-axis, is larger in comparison to the values for CuCN and AgCN. This also suggests that nature of boding in AuCN is stronger in comparison to CuCN and AgCN.

**E. Thermal Expansion Behavior**

The lattice parameter as a function of temperature has been reported from neutron diffraction measurements at temperature ranging from 90 K to 450K[39]. The measurements show that the *c* lattice parameter decreases with increase in temperature, however, lattice parameter *a* (=*b)* shows positive expansion behaviour. The overall volume thermal expansion is found to be positive in all three cyanides and has similar magnitude. The negative thermal expansion behavior along c axis is largest in CuCN and least in AuCN. So also, the positive expansion along a and b-axis is largest in CuCN and least in AuCN. We have computed the thermal expansion behavior using the quasiharmonic approximation. Each phonon mode of energy $E_i$ contributes to the volume thermal expansion coefficient[58] given by $\alpha_V = \frac{1}{BV}\sum_i \Gamma_i C_{Vi}(T)$, where $V$ is the unit cell volume, $B$ is the bulk modulus, $\Gamma_i$ ( = $-\partial \ln E_i/\partial \ln V$) are the mode Grüneisen parameters and $C_{Vi}$ the specific-heat contributions of the phonons in state i (= $\mathbf{q}j$) of energy $E_i$. The volume dependence of phonon frequency is used to calculate the thermal expansion behavior. The phonon spectra in the entire Brillouin zone have been calculated at two volume corresponding to the experimental structures at 10 K and 310 K. The calculated Grüneisen parameters are shown in Fig.8(a). It can be seen that low energy modes below 4 meV have large positive Grüneisen parameters. The calculated partial density of states shows that the contribution at such low energies is mainly from the M( =Cu, Ag and Au) atoms. As shown in Fig. 5 the calculated transverse acoustic modes in HT-CuCN and AgCN and AuCN are unstable. For the thermal expansion calculation the phonon energies have been calculated at 8000 q points (72000 phonon modes) in the entire Brillouin zone. We find that among these 72000 modes, the number of unstable modes in HT-CuCN and AgCN and AuCN are 558, 453 and 75 respectively, which is less than 1%. The temperature dependence of the unit cell volume is calculated without including the unstable modes. The calculations are qualitatively in good agreement with the observed thermal expansion behavior in AuCN, but underestimated the experimental magnitude in other two compounds (Fig 9). The underestimate might be related to the C/N disorder as discussed below.

The experimental[39, 44] value of the coefficient of negative thermal expansion (NTE) along the chain direction ($\alpha_c$) for HT-CuCN, AgCN and AuCN is $-27.9\times10^{-6}$ K$^{-1}$, $-14.8\times10^{-6}$ K$^{-1}$ and $-6.9\times10^{-6}$ K$^{-1}$ respectively, while positive thermal expansion (PTE) in the a-b plane ($\alpha_a$) is $74.8\times10^{-6}$ K$^{-1}$, $65.7\times10^{-6}$ K$^{-1}$ and $57.4 \times10^{-6}$ K$^{-1}$ respectively. As noted above, among the three compounds HT-CuCN has the highest C/N disorder and it has also the highest positive as well as negative thermal expansion coefficients. AuCN has the least C/N disorder and has the smallest values of NTE and PTE coefficients. As mentioned above, the ab-initio calculations performed with the ordered structures exhibit the highest number of unstable modes for HT-CuCN, while AuCN show the least number of unstable modes. It seems C/N disorder stabilizes the structure. Among cyanides, nickel cyanide Ni(CN)$_2$ has a long-range ordered structure in two dimensions (*a-b* plane) but a high degree of stacking disorder in the third dimension. The compound exhibits[59] NTE in two dimensions ($\alpha_a= -7 \times 10^{-6}$ K$^{-1}$) along with a very large PTE coefficient ($\alpha_c = 61.8 \times 10^{-6}$ K$^{-1}$) perpendicular to the layers. Here again it can be seen that disorder along c-axis results in large overall volume thermal expansion ($\alpha_V = 48.5 \times 10^{-6}$ K$^{-1}$). It appears that the C/N disorder contributes towards positive thermal expansion behavior. The order-disorder transitions at around 400 K in ZrW$_2$O$_8$ reduces[21] the overall NTE coefficient.

As noted above, the linear thermal expansion coefficient along 'a-' and 'c-' axis are found to be positive and negative respectively. We are interested to find the modes which have large negative and positive Grüneisen parameters and contribute towards thermal expansion behavior. The estimated Grüneisen parameters ($\Gamma_i$) and the specific-heat contribution of modes ($C_{Vi}$) from the ab-*initio* calculations have been used to estimate the contribution of the various phonons to the thermal expansion (Fig. 8(b)) as a function of phonon energy at 300 K. The maximum contribution to $\alpha_V$ seems to be from the low-energy modes below 10 meV. The calculated volume dependence of phonon dispersion curves for HT-CuCN, AgCN and AuCN are shown in Fig. 5. The displacement pattern of a few zone boundary phonon modes, has been plotted (Fig. 10). The mode assignments, phonon energies and Grüneisen parameters are given in the figures. As mentioned above, HT-CuCN and AgCN crystallize in the same space group (R3m), hence the eigen vector pattern for symmetrically equivalent phonon modes would be similar. The investigation of the displacement pattern of the eigenvectors shows that the phonon modes have mainly two kinds of dynamics. One which involves atom vibration along the chain and the other in which atoms vibrate perpendicular to the chain.

For HT-CuCN and AgCN, the adjacent -M-C≡N-M- chains are shifted by ±c/3 along c-axis. We find that lowest zone-boundary modes at F and LD points in the Brillouin zone are found to be unstable. For the LD point mode (Fig. 10), within a chain, the M and C≡N move with equal displacements. The movement of atoms in the adjacent chains is found to be out-of-phase with each other. The motion of the atoms in F-point mode is similar to that in LD-point mode. However for F-point mode there is a small component of displacement in the a-b plane. Both the modes are found to become more unstable on compression of the lattice. Such type of modes would contribute maximum to the NTE along c-axis. However in case of AuCN, the K point mode (Fig. 10) also shows sliding of -M-C≡N-M- chains out-of-phase with each other. The mode is found to have small positive $\Gamma$ of 1.1. It seems that the chain sliding modes mainly contribute to negative $\alpha_c$ in HT-CuCN and AgCN compounds and this contribution is not seem in AuCN.

The vibrational amplitude along the chain would depend on the nature of bonding between metal and cyanide (-C≡N-) as well as on the atomic mass of metal ion. As mentioned above, this bonding in HT-CuCN and AgCN seems to be similar. The smaller mass of Cu (63.54 amu) would lead to large amplitude of thermal vibration along the chain in comparison to Ag (107.87 amu) compound, which indicates that the contraction along the -M-C≡N-M- chain would be more in the HT-CuCN in comparison to the AgCN, which is qualitatively in agreement[39] with the observed NTE behavior in these compounds. Several modes in which the atoms move perpendicular to the chain have positive Grüneisen parameters and would be responsible for positive thermal expansion behavior.

## V. CONCLUSIONS

We report a comparative study of the dynamics of quasi one-dimensional metal cyanides MCN (M=Cu, Ag and Au) using inelastic neutron scattering measurements as well as through first principle calculations based on the density functional theory. A good match between the calculated phonon density of states and that derived from inelastic neutron scattering measurements is obtained. The calculated thermal expansion behavior is found to be in qualitative agreement with the available experimental data. We have also identified the specific nature of phonon modes associated with the sliding of -M-C≡N-M- chains along the hexagonal axis and vibrations in the hexagonal plane, which are responsible for the anomalous thermal expansion behavior in these cyanides. The nature of the chemical bonding is found to be similar in HT-CuCN and AgCN, which is significantly different from that in AuCN. The computed elastic constants and Born effective charges are correlated with the difference in nature of bonding in metal cyanides.

TABLE I: The structure of various cyanides[42,56] (T=10 K) as used in the ab-initio calculations of phonon spectra. The 'a' and 'c' lattice constants and atom coordinates in the hexagonal unit cell are given.

|  | HT-CuCN(R3m) | AgCN(R3m) | AuCN( P6mm) |
|---|---|---|---|
| a(Å) | 5.912 | 5.905 | 3.343 |
| c(Å) | 4.849 | 5.291 | 5.098 |
| V( Å$^3$)/Z | 49.407 | 53.481 | 49.828 |
| M (Cu, Ag Au) | 1/3,2/3,1/3 | 1/3,2/3,1/3 | 0,0,0 |
| C | 1/3,2/3,0.714 | 1/3,2/3,0.724 | 0,0,0.387 |
| N | 1/3,2/3,0.952 | 1/3,2/3,0.942 | 0,0,0.613 |

TABLE II: The various elastic constants of metal cyanides MCN (M=C, Ag and Au) in unit of GPa at T=0 K.

|  | $C_{11}$ | $C_{33}$ | $C_{44}$ | $C_{66}$ | $C_{12}$ | $C_{13}$ |
|---|---|---|---|---|---|---|
| HT-CuCN | 14.0 | 536.0 | 4.0 | 0.4 | 6.1 | 11.0 |
| AgCN | 18.5 | 387.0 | 5.2 | -0.3 | 8.1 | 16.0 |
| AuCN | 28.4 | 755.3 | 6.5 | 2.2 | 15.3 | 14.0 |

TABLE III: The Born effective charges of various atoms in unit of *e*. ($Z_{yy}=Z_{xx}$; $Z_{xy}=Z_{xz}=Z_{yx}=0$)

| **Atom** | $Z_{xx}$ | $Z_{zz}$ |
|---|---|---|
| C(HT-CuCN/AgCN/AuCN) | -0.3/-0.4/-0.5 | 1.4/0.8/1.5 |
| N(HT-CuCN/AgCN/AuCN) | -0.6/-0.6/-0.5 | -1.1/-1.2/-1.5 |
| Cu/Ag/Au | 0.9/1.0/1.0 | -0.3/0.4/0.0 |

TABLE IV: The various bond length in metal cyanides MCN (M=C, Ag and Au) in unit of Å.

| Bond length | HT-CuCN | AgCN | AuCN |
|---|---|---|---|
| C-N | 1.174 | 1.169 | 1.163 |
| C-M | 1.841 | 2.040 | 1.960 |
| N-M | 1.835 | 2.081 | 1.976 |

FIG. 1 (Color online) The structure of AuCN and HT-CuCN/AgCN as used in the ab-intio calculations. Key: C, red sphere; N, blue sphere; Cu/Ag/Au green sphere

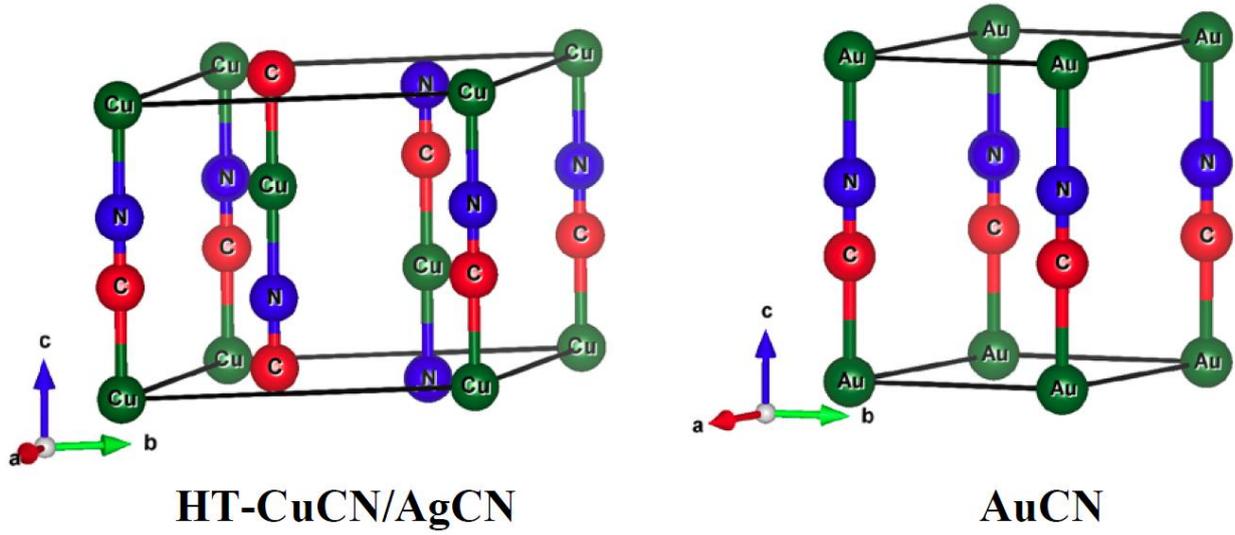

**HT-CuCN/AgCN**  **AuCN**

FIG. 2 (Color online) The measure neutron inelastic spectra MCN (M=Cu, Ag and Au) at 150 K, 240 K and 310 K

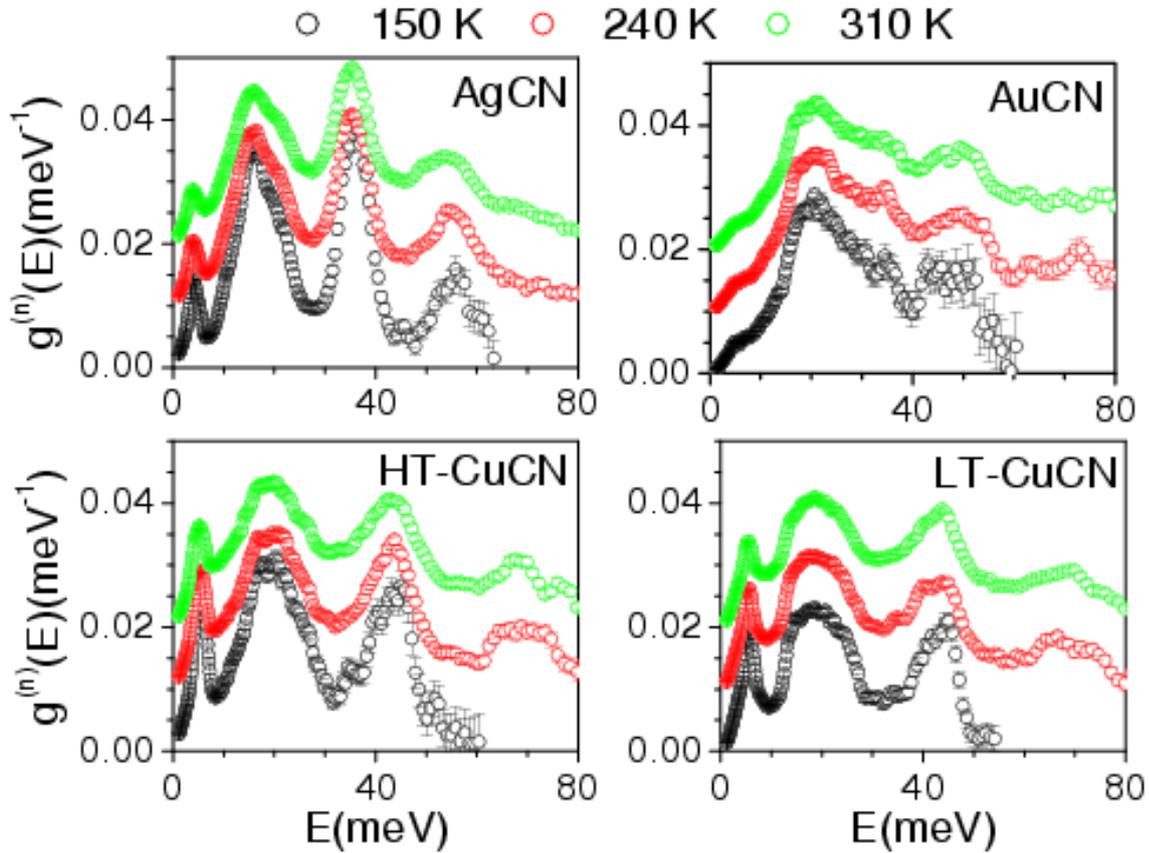

FIG. 3 (Color online) (a) Comparison of the experimental phonon spectra for MCN (M=Cu, Ag and Au) at 310 K. (b) Comparison of the experimental phonon spectra for LT and HT phases of CuCN at 150 K.

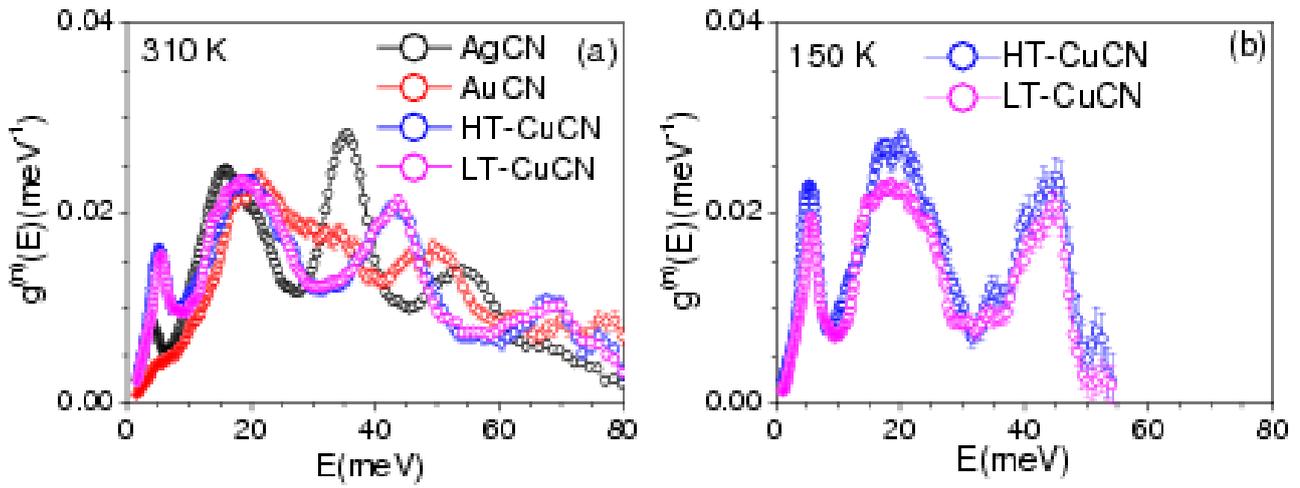

FIG. 4 (Color online) The comparison between the measured (310 K) and calculated phonon spectra of MCN (M=Cu, Ag and Au).

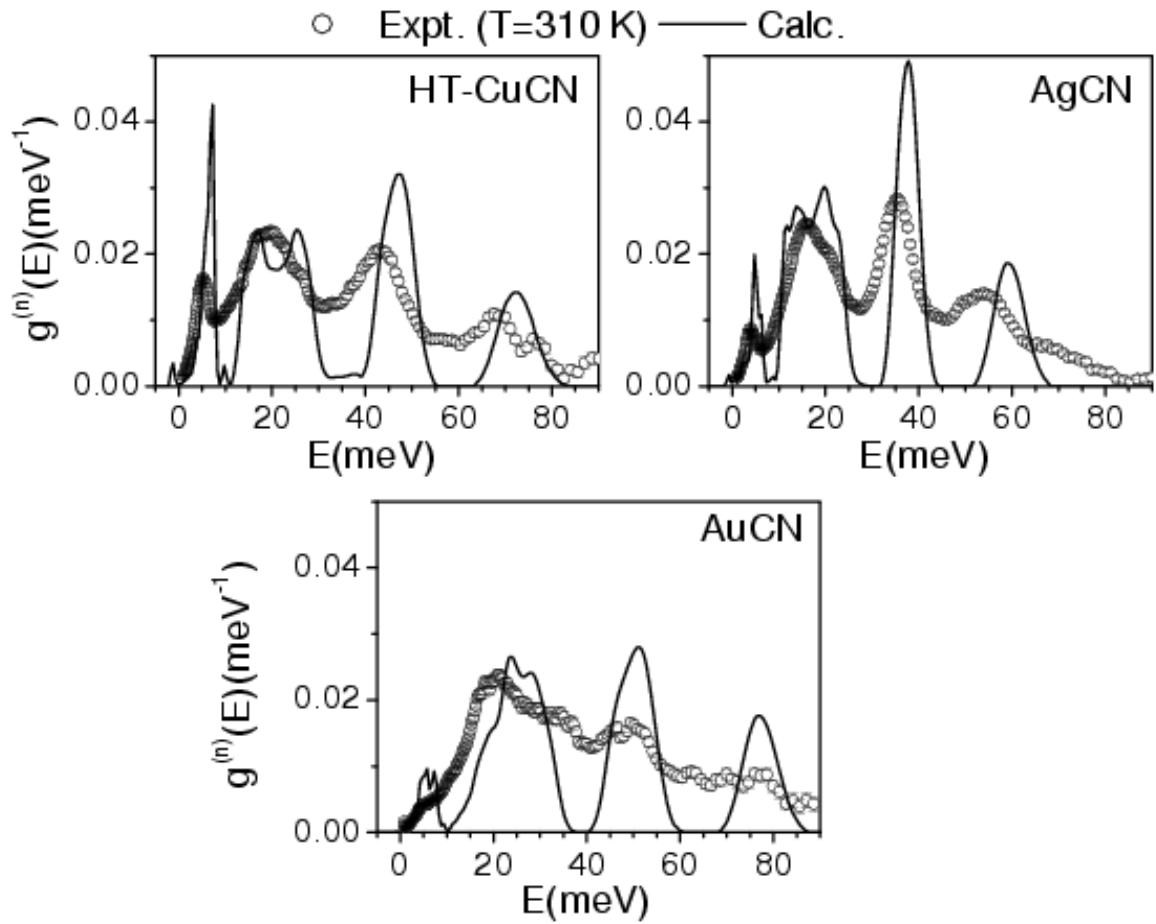

FIG. 5 (Color online) The calculated dispersion relation along various high symmetry direction of MCN (M=Cu, Ag and Au) at lattice constant at 10 K (black) and 310 K (red). The C-N stretching modes at about 270 meV are not shown. The Bradley-Cracknell notation is used for the high-symmetry points. HT-CuCN/AgCN: T1(1/2,1/2,-1/2)$_R$ ≡ (0, 1, 1/2)$_H$ , Γ(0,0,0)$_R$ ≡ (0, 0, 0)$_H$ ,T2(1/2,1/2,1/2)$_R$ ≡ (0, 0, 3/2)$_H$, F(1/2,1/2,0)$_R$ ≡ (0, 1/2, 1)$_H$, L(0,1/2,0)$_R$ ≡ (-1/2,1/2,1/2)$_H$; AuCN: Γ(0,0,0)$_H$, A(0 0 1/2)$_H$, K(1/3,1/3,0)$_H$, H(1/3 1/3 1/2)$_H$, L(1/2 0 1/2)$_H$ and M(1/2,1/2,0)$_H$.   Subscript R and H correspond to rhombohedral and hexagonal notation respectively.

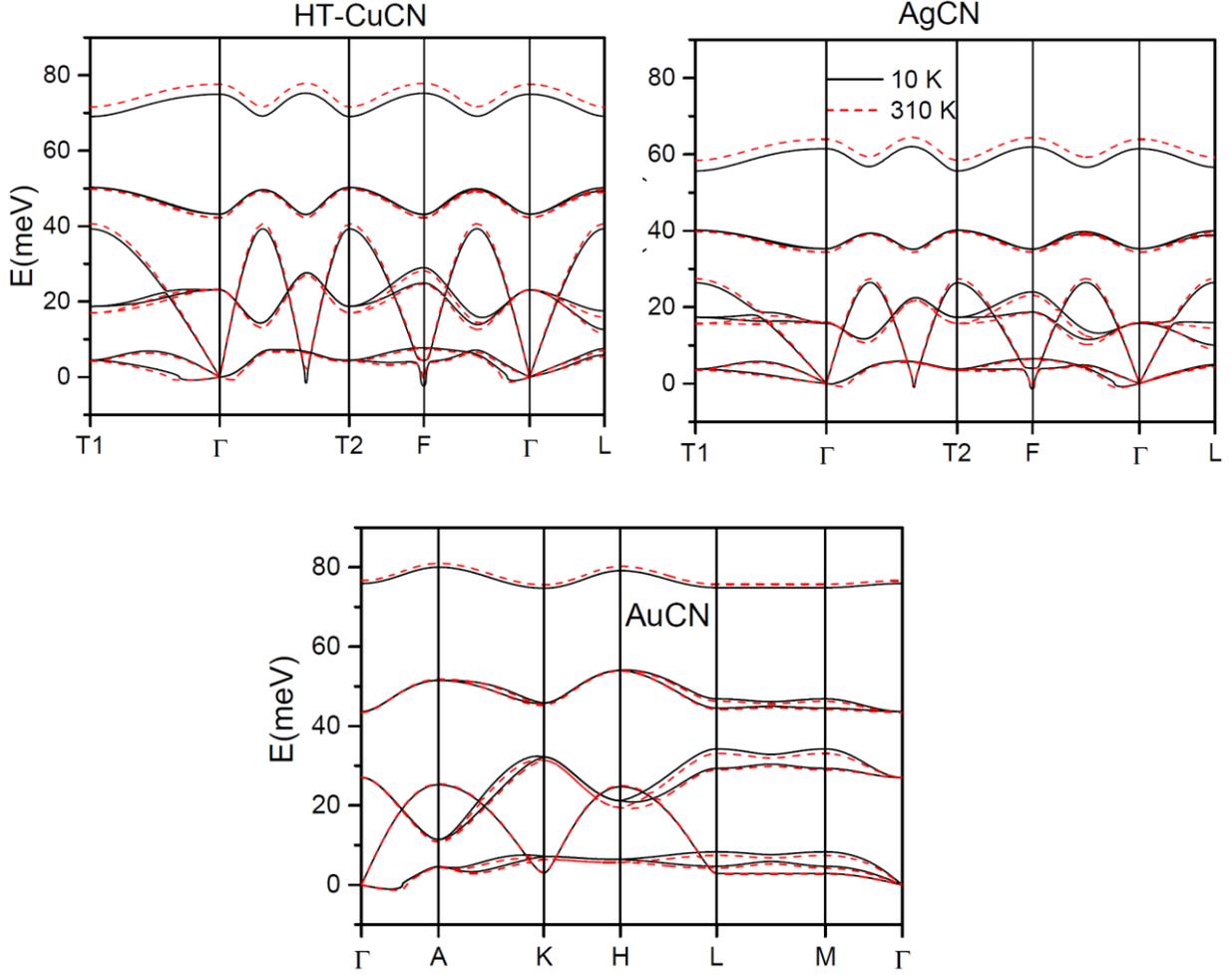

FIG. 6 (Color online) The calculated phonon partial density of states of various atoms in MCN (M=Cu, Ag and Au) for structure at 10 K. The *x*-scale the C-N stretching modes at about 270 meV are not shown.

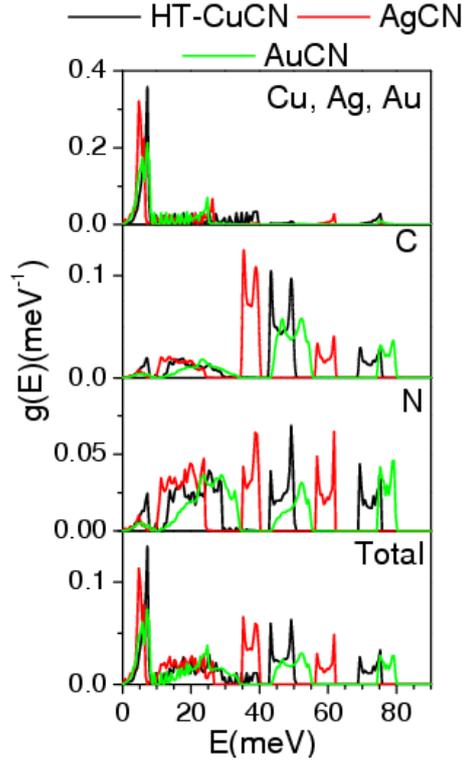

FIG. 7 (Color online) The partial phonon density of states of MCN (M=Cu, Ag and Au) at structural parameters corresponding to 10 K (black) and 310 K (red). The *x*-scale the cyanide stretching modes at about 270 meV are not shown.

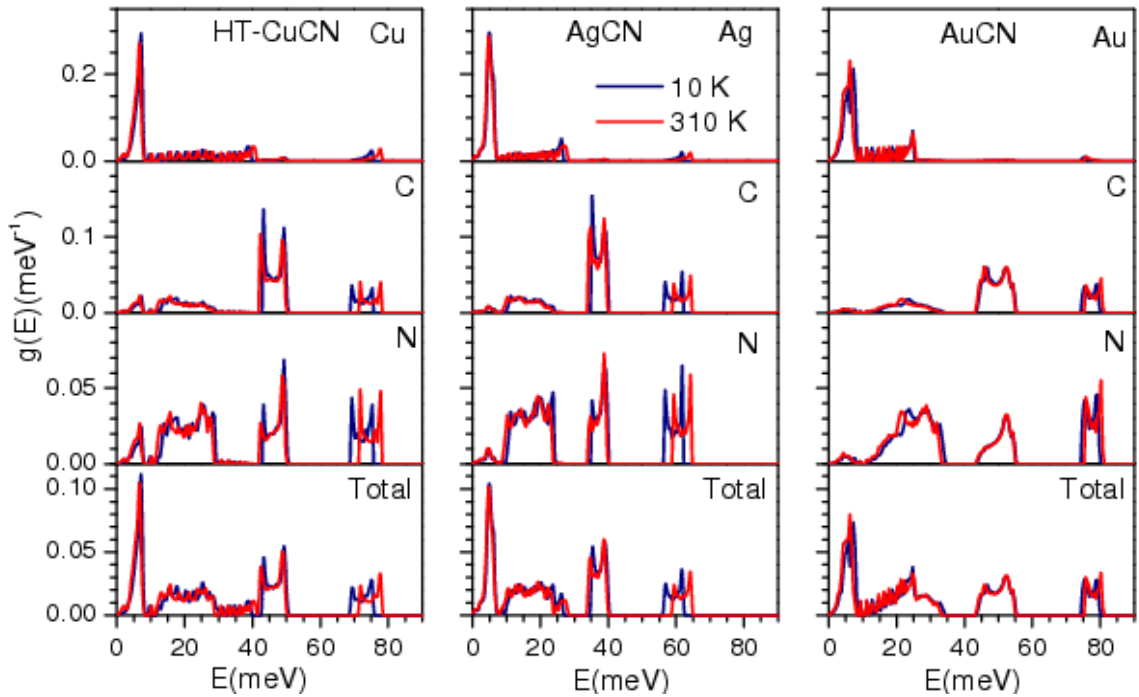

FIG. 8 (Color online) (a) The calculated average Grüneisen parameters Γ(E) averaged over various phonon of energy $E$ in the whole Brillouin zone. (b) The contribution of phonons of energy $E$ to the volume thermal expansion coefficient ($\alpha_V$) as a function of $E$ at 300 K.

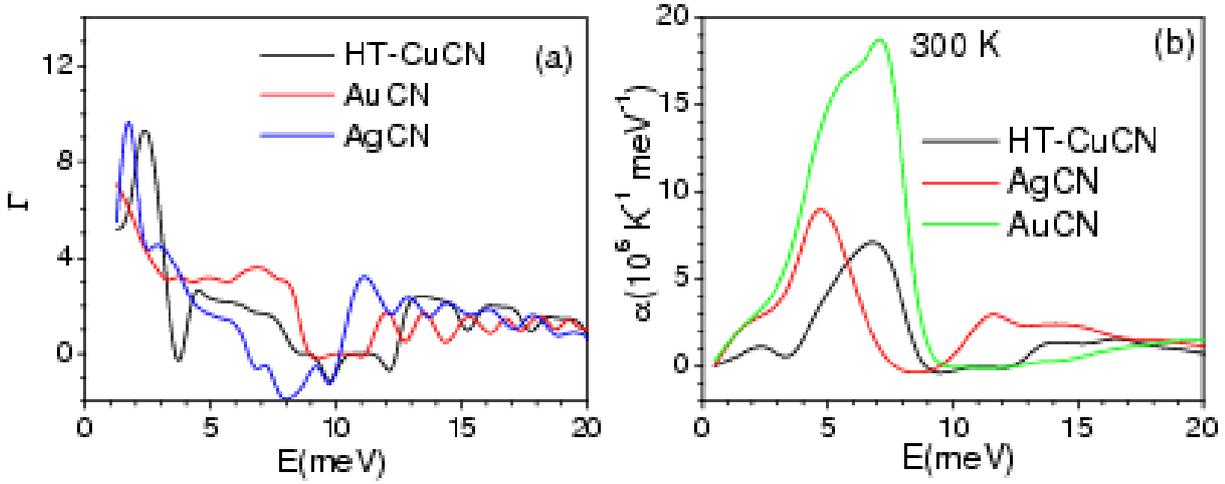

FIG. 9 (Color online) The calculated and experimental thermal expansion behavior of MCN (M=Cu, Ag and Au).

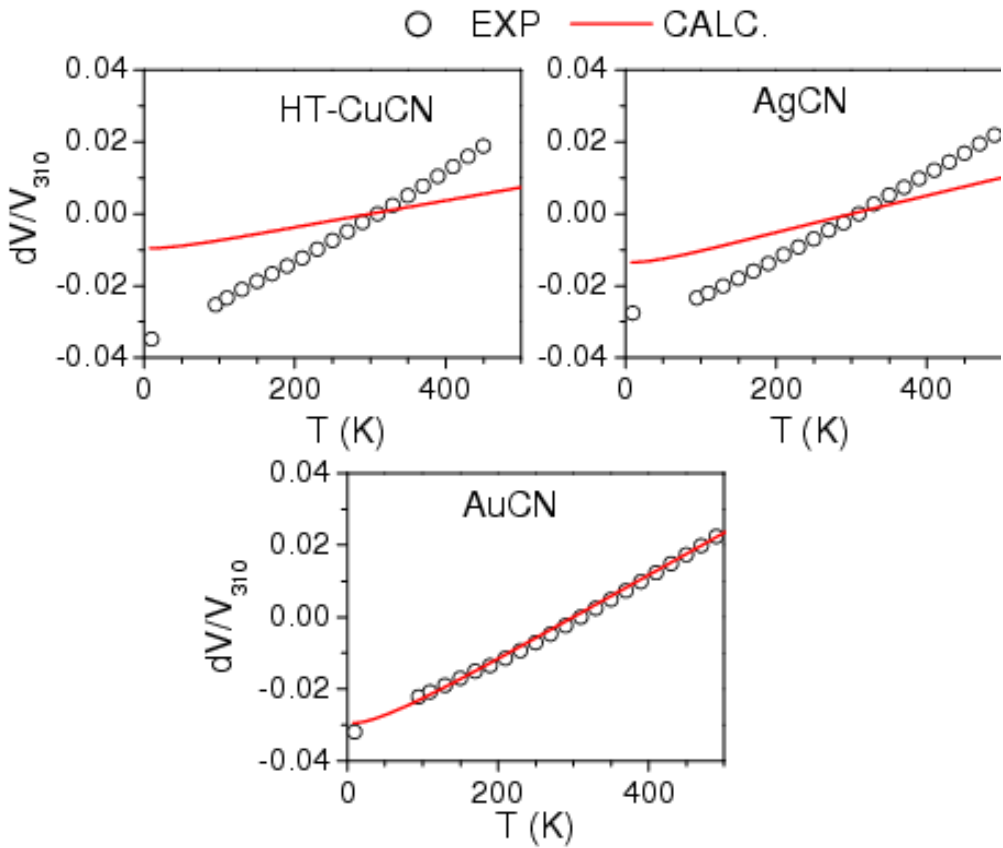

FIG. 10 (Color online) The calculated displacement pattern of various phonon modes in AuCN and HT-CuCN and corresponding Grüneisen parameters. The first line below each figure represents the size of the supercell. The second line below the figure give the high symmetry point, phonon energies and Grüneisen parameters, respectively. In the bottom panel (HT-CuCN and AgCN) the second and third line below the figure corresponds to HT-CuCN and AgCN respectively. The Bradley-Cracknell notation is used for the high-symmetry points. AuCN: A=(0 0 1/2)$_H$, K(1/3,1/3,0)$_H$ and M(1/2,1/2,0)$_H$; HT-CuCN/AgCN: L(0,1/2,0)$_R$ ≡ (-1/2,1/2,1/2)$_H$, T1, T2 (1/2,1/2,-1/2)$_R$ ≡ (0, 1, 1/2)$_H$, F(1/2,1/2,0)$_R$≡(0,0.5,1)$_H$, LD(-2/3,1/3,1/3)$_R$ ≡ (1,0,0)$_H$. Subscript R and H correspond to rhombohedral and hexagonal notation respectively. The c-axis is along the chain direction, while a and b-axis are in the horizontal plane. Key: C, red sphere; N, blue sphere; Cu/Ag/Au green sphere.

*The Grüneisen parameters values of unstable F and LD-point modes are not given. The modes are found to become more unstable on further compression of the lattice. Such type of modes would contribute maximum to the NTE along c-axis.

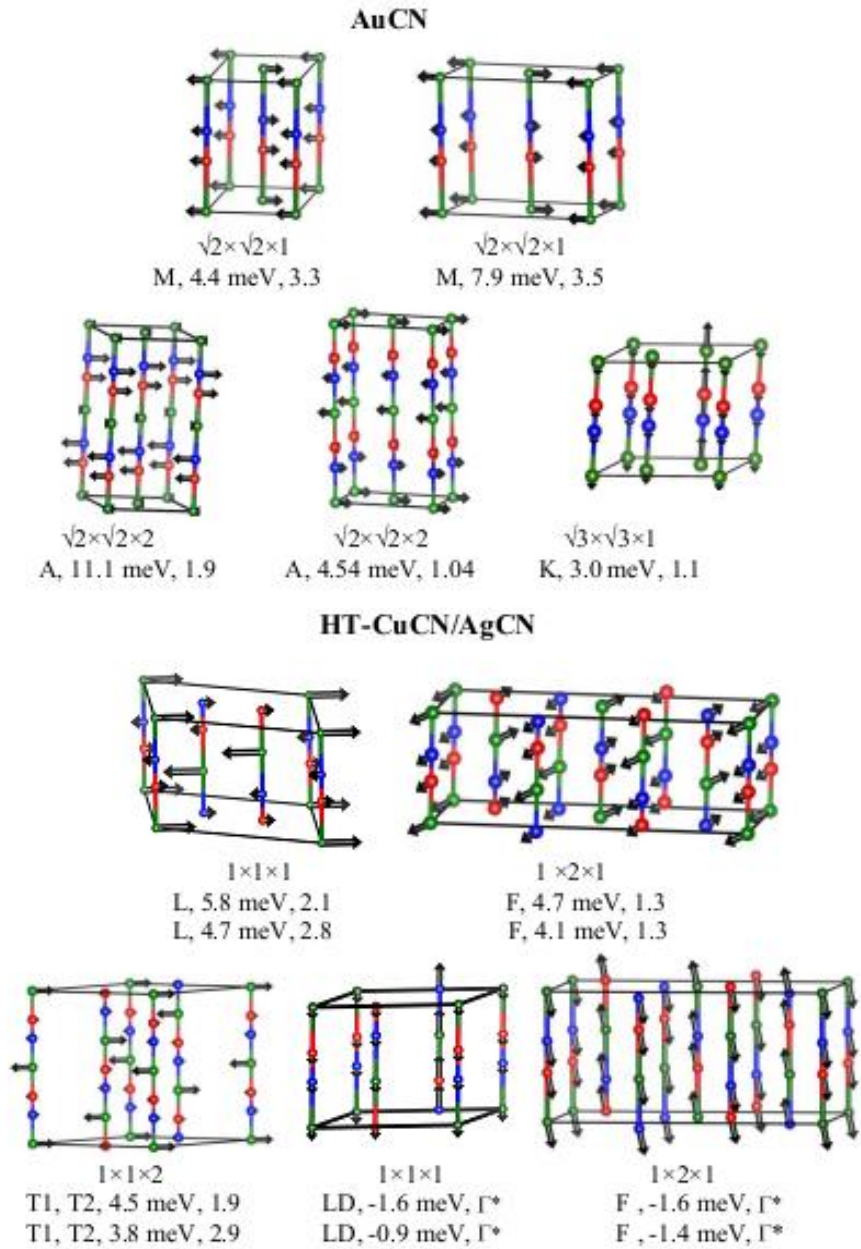